\documentclass[twocolumn,showpacs,preprintnumbers,amsmath,amssymb,dvipdfmx]{revtex4-1}

\usepackage{graphicx}
\usepackage{dcolumn}
\usepackage{bm}
\usepackage{color}
\usepackage{ulem}
\usepackage[colorlinks=true,citecolor=blue,linkcolor=blue,filecolor=blue,urlcolor=blue]{hyperref} 

\begin{document}

\title{
Charge glass from supercooling topological-ordered liquid
}%

\author{Kouki Kimata$^1$, Harukuni Ikeda$^{1,2}$, and Masafumi Udagawa$^1$}%
\affiliation{%
$^1$Department of Physics, Gakushuin University, Mejiro, Toshima-ku, Tokyo 171-8588, Japan\\
$^2$Yukawa Institute for Theoretical Physics, Kyoto University, Kyoto 606-8502, Japan
}%

\date{\today}

\begin{abstract}
Topological order characterizes a class of quantum and classical many-body liquid states that escape the conventional classification by spontaneous symmetry breaking. Many properties of the topological-ordered states still await a clear understanding, and nature of phase transition dynamics is one of them. Normally, when a liquid freezes into a solid, crystallization starts with nucleation and a solid domain quickly grows on the surface of the expanding nucleus, and the domains evolve into macroscopic size. In this work, we reveal that the crystallization of the topological-ordered liquid proceeds in a fundamentally different way. The topological-ordered phase is characterized by a global conserved quantity and its conjugate fractional charge, which we call a flux and a triplet in our working system of the charge Ising model on a triangular lattice. In contrast to the normal crystallization process, the phase transition is driven by the diffusive motion of triplets, which is required to change the value of conserved fluxes to exit the topological-ordered phase. In order to complete crystallization, triplets must spend a divergently long time to diffuse over a macroscopic distance across the system, which results in glassy behavior. Reflecting the diffusive motion of triplets, the initial crystallization process shows slowing down with unusually small Avrami exponent $\sim0.5$. These anomalous dynamics are specific to the crystallization from topological-ordered liquid, and well account for the main features of charge glass behavior exhibited by the organic conductors, $\theta$-(BEDT-TTF)$_2$X(SCN)$_4$. 
\end{abstract}

\maketitle
{\it Introduction}:  
The origin of glass is a classic problem that has continuously stimulated interest in various fields of science~\cite{debenedetti2001,berthier2011}.
Among possible origins of glass, geometrical frustration is sometimes raised as a main suspect~\cite{ANDERSON1978291,Kivelson1995,Tarjus2005,schmalian2000,westfahl2001}. According to a popular scenario, geometrical frustration brings about strong phase competition between macroscopically degenerate low-energy states and thereby considerably delays the ordering dynamics. While this idea may sound reasonable, it still remains hypothetical. It is strongly desired to clarify the role of geometrical frustration in glassy dynamics.

A class of organic compounds, $\theta$-(BEDT-TTF)$_2$X(SCN)$_4$~\cite{PhysRevB.57.12023} is one of the typical systems where geometrical frustration is suspected of its glassy behavior~\cite{Kagawa:2013aa}. The lattice structure of this compound is an alternating stack of conductive BEDT-TTF molecule layers and insulating anion X(SCN)$_4$ layers. On the BEDT-TTF layer, one hole exists per molecule on average, and accounts for the conduction of this system. The BEDT-TTF molecules are placed on a triangular lattice, subject to strong geometrical frustration. The glassy behavior of the system is controlled by the species of anion, X(SCN)$_4$. For X$=$RbZn, the system exhibits metal-insulator transition at $T_{\rm c}=195$K, accompanied with charge ordering of horizontal-stripe-type~\cite{doi:10.1143/JPSJ.73.116,PhysRevB.62.R7679,PhysRevB.65.085110}. A compound with X$=$TlZn ($\theta_m$-TlZn) takes a monoclinic crystal structure, and this system shows charge ordering at $T_{\rm c}=170$K with diagonal-stripe-type charge order~\cite{doi:10.1126/science.aal3120}. 
The compound with X$=$CsZn is exceptional: it stays disordered down to low temperatures~\cite{doi:10.1143/JPSJ.74.2631,PhysRevB.77.115113,Nad_2008}, with weak anomalies observed around 100K~\cite{PhysRevB.89.121102}, which might be connected to the structural instabilities~\cite{PhysRevB.105.L041114,PhysRevResearch.6.023003,PhysRevResearch.5.013024}.

Remarkably, even for the systems showing the tendency of ordering, X$=$RbZn and TlZn systems, the charge ordering behavior depends on the cooling rate: Charge order occurs only when the system is cooled down slowly below $T_{\rm c}$.
If the system is cooled rapidly enough, the charge ordering is suppressed~\cite{Kagawa:2013aa,doi:10.1126/science.aal3120,doi:10.1126/science.aal2426,PhysRevB.105.205111}, as is the case with X$=$CsZn.
The state with suspended charge order is called charge glass, and many theoretical analyses have been conducted for its clarification, focusing on various aspects of the system, such as geometrical frustration, the long-range nature of Coulomb interaction, and the strong phase competition~\cite{cryst2031155,doi:10.1143/JPSJ.75.123704,Hotta_2011,PhysRevB.79.195124,PhysRevB.74.193107,PhysRevB.103.L201106,PhysRevLett.115.025701,doi:10.7566/JPSJ.89.034003,doi:10.1143/JPSJ.81.063003,PhysRevLett.98.206405,ydc4-cfrm}.

\begin{figure}[h]
\begin{center}
\includegraphics[width=0.5\textwidth]{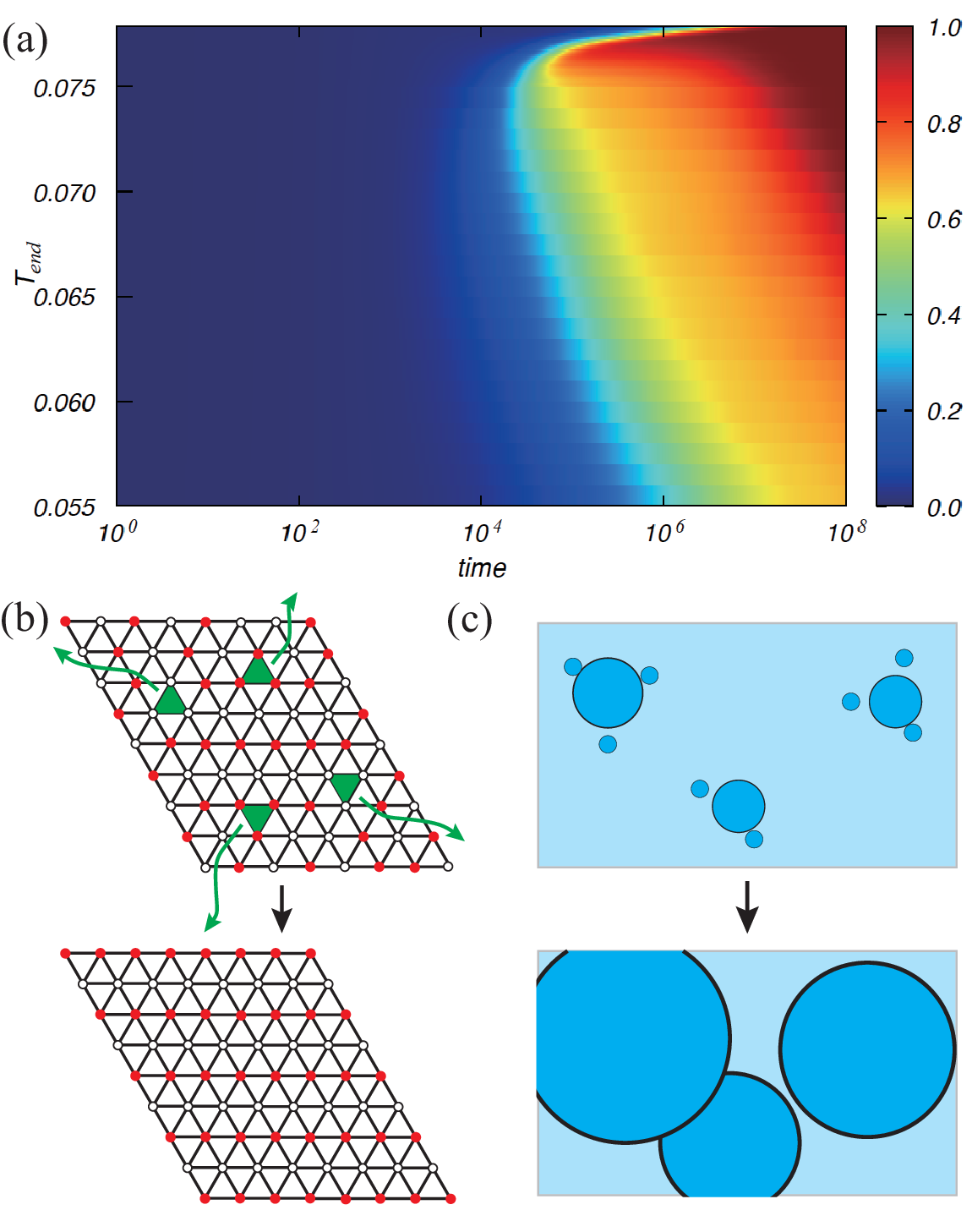}
\end{center}
\caption{\label{Fig0} 
(Color online) (a) The numerically obtained Time-Temperature-Transformation (TTT) diagram showing the time evolution of the ordered fraction, $\Delta_{\rm CO}$, after the system is quenched to the temperature, $T_{\rm end}$, in the zero-triplet quench protocol. 
(b) Schematic illustration of crystallization by diffusion: triplets, represented by upward and downward triangles, diffuse through the system and induce charge ordering.
(c) Schematic illustration of conventional crystallization, where small ordered domains grow at the surfaces of existing ones and expand into macroscopic regions.
}
\end{figure}

Experimental studies on charge glass have recently made great advances. Through the resistivity~\cite{doi:10.1126/science.aal2426,doi:10.1126/science.aal3120}, NMR~\cite{doi:10.1126/science.aal2426}, and Raman spectroscopy~\cite{Murase:2023aa},  the spatial fraction of the charge-ordered domain is mapped out in the time-temperature-transformation (TTT) diagram, enabling complete resolution of ordering dynamics with respect to time and temperature axes. The TTT diagram has several remarkable features.  Firstly, the charge order is developed only in the quench to the vicinity of $T_{\rm c}$ in an experimentally observable time range. Secondly, the ordering time, $t_{\rm CO}$, shows non-monotonic temperature dependence, and exhibits a minimal value at the nose temperature, $T_{\rm nose}$, which is estimated around 160 K for both RbZn and TlZn compounds, at which the qualitative change of ordering dynamics has been reported~\cite{doi:10.1126/science.aal2426,Murase:2023aa}.

The initial ordering process also exhibits a notable feature: in the TlZn system, the ordering proceeds quite slowly. The ordering dynamics is well fitted by Johnson-Mehl-Avrami-Kolmogolov (JMAK) curve, $f(t) = 1-e^{-Kt^n}$, throughout the entire charge-ordering process~\cite{doi:10.1126/science.aal3120}, where the Avrami exponent $n$ characterizes the acceleration rate of initial dynamics; $f(t)\sim Kt^n$ for $t\to 0$. Typically, the ordering process accelerates because ordered domains predominantly grow at the expanding surfaces of already developed domains, and the exponent $n$ is related to the spatial dimension $d$, as $n=d+1$, taking a value larger than 1. In contrast, for the TlZn system, an unusually small exponent $n\sim0.56$ has been reported to fit the ordering curve~\cite{doi:10.1126/science.aal3120}, implying that the ordering process once slows down after it begins. For the RbZn system, a slower ordering process has also been reported below $T_{\rm nose}$, although the Avrami exponent takes a more conventional value, $n=2-3$~\cite{doi:10.1126/science.aal2426}.
 
\begin{figure*}[t]
\begin{center}
\includegraphics[width=0.83\textwidth]{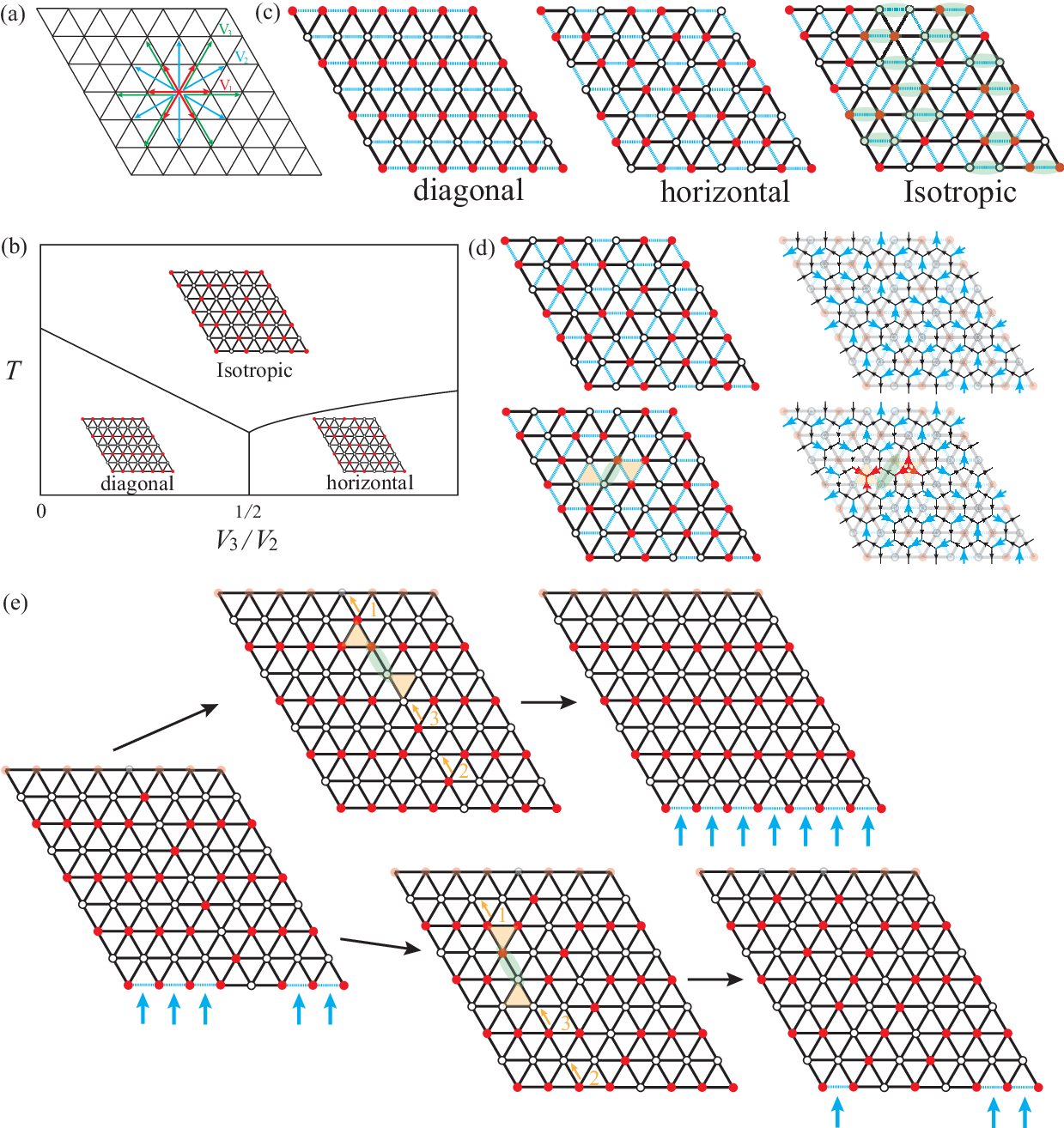}
\end{center}
\caption{\label{Fig1} 
(Color online) Model and topological constraint of the charge Ising model on a triangular lattice. (a) Geometry of the lattice and definition of the nearest, second, and third neighbor interactions $V_{1}$, $V_{2}$, and $V_{3}$.
(b) Schematic phase diagram of Hamiltonian, Eq.~(\ref{eq:Hamiltonian}). At zero temperature, diagonal and horizontal charge-ordered phases become ground states for $V_3<\frac{V_2}{2}$ and $V_3>\frac{V_2}{2}$, respectively. At higher temperatures, equilibrium states are described mostly by the states satisfying the Coulomb rule, provided $V_2$ and $V_3$ are sufficiently smaller than $V_1$.
(c) Typical charge configurations. Dashed blue lines show frustrated bonds. For the isotropic configuration, the horizontal frustrated bonds are highlighted with green ovals. Each row supports the same number of frustrated bonds: $L_1=2$.
(d) Mapping to the dual honeycomb lattice, showing (top) a Coulomb-phase configuration and (bottom) a pair of triplets created by shifting a single charge along a bond indicated by a green oval. Orange triangles show triplet positions.
(e) Illustration of triplet propagation: a pair of triplets are created, diffuse along orange arrows, 1, 2, 3, and annihilate after circulating the system, thereby changing the global flux sector. The top (bottom) panel shows the propagation of upward (downward) triangle, which has negative (positive) charge in the arrow representation. Accordingly, the flux value $L_1$  increases (decreases) by 2, as indicated by blue arrows.
}
\end{figure*}

Despite the intensive theoretical studies so far, the dynamical aspects of charge glass remain elusive. In this work, we focus on the structure of the disordered phase above $T_{\rm c}$. Due to the geometrical frustration, this disordered phase naturally develops a spatial pattern similar to the so-called Coulomb phase, which typically appears in frustrated magnetic systems and is often referred to as a (classical) spin liquid~\cite{annurev:/content/journals/10.1146/annurev-conmatphys-070909-104138,Udagawa2021}.

This Coulomb phase above $T_{\rm c}$ is characterized by topological order: there appears a global conserved quantity, which we call as a flux, and its conjugate fractional charge, referred to as a triplet. These two objects are closely connected: to alter the value of conserved flux, triplets must be pair-created, transported across the system, and pair-annihilated. The high-temperature Coulomb phase belongs to the isotropic flux sector, while the low-temperature charge-ordered phases belong to the anisotropic sector. Consequently, when the system is suddenly cooled to the ordered phase, triplets must diffuse across macroscopic distances to change the flux sector and develop charge orders. This necessity of triplet motion introduces a kinetic bottleneck and results in glassy behavior.

Based on this picture, we could indeed reproduce the TTT diagram with the nose temperature, as experimentally observed  [Fig.~\ref{Fig0} (a)]. The glassy behavior rooted in topological order indicates a new type of phase-transition dynamics, which we term ``crystallization by diffusion" [Fig.~\ref{Fig0} (b)]. In this mechanism, ordering proceeds by the diffusion of triplets over macroscopic distances, resulting in the macroscopically long relaxation time. Moreover, the diffusive nature of triplet motion results in a slow onset of ordering dynamics, which leads to the smaller Avrami exponent. These features are in sharp contrast to the normal crystallization process, where the nucleated ordered domains quickly expand to cover the whole system [Fig.~\ref{Fig0} (c)]. In the following, we adopt the charge Ising model as the simplest microscopic model to embody this mechanism, which reproduces the main features of charge glass observed in $\theta$-(BEDT-TTF)$_2$X(SCN)$_4$.

{\it Model}: 
As a model to address the charge glass, we start with the charge Ising model defined on a triangular lattice,
\begin{eqnarray}
\mathcal{H} = V_1\sum_{\langle i,j\rangle_1}n_in_j + V_2\sum_{\langle i,j\rangle_2}n_in_j + V_3\sum_{\langle i,j\rangle_3}n_in_j.
\label{eq:Hamiltonian}
\end{eqnarray}
We consider a triangular lattice with the total number of sites $N=L^2$, imposing periodic boundary conditions in both directions [Fig.~\ref{Fig1}(a)]. In Eq.~(\ref{eq:Hamiltonian}), $n_i$ denotes the conserved charge variable, corresponding to the presence or absence of charge at site $i$, and takes $n_i=1$ (charge) or $0$ (vacancy). To represent the quarter-filled BEDT-TTF layer (half-filled in the spinless case), we fix the total number of charges to be $N_{\rm e}\equiv\sum_in_i = N/2$. Reflecting the long-range nature of the Coulomb interaction, we include couplings up to the third-nearest neighbors, $V_1$, $V_2$, and $V_3$, as illustrated in Fig.~\ref{Fig1} (a). All couplings are positive, and we set $V_1=1$ as a unit of energy.

If $V_1$ is dominant over $V_2$, and $V_3$, the ground state is the diagonal (horizontal) charge ordered state for $V_3<\frac{V_2}{2}$ ($V_3>\frac{V_2}{2}$), as shown in the schematic phase diagram [Fig.~\ref{Fig1}(b)]. The phase diagram was obtained numerically in the $V_1\to\infty$ limit, and the transition to the diagonal phase is found to be of first order~\cite{PhysRevLett.116.197201}. The diagonal and horizontal order are the same as the charge ordering patterns observed for X$=$TlZn and RbZn systems, justifying the model Eq.~(\ref{eq:Hamiltonian}) as a minimal model to describe the charge ordering in these compounds.

{\it Flux and triplet}: 
The key feature of this model is a topological constraint that strongly affects its dynamics. It can be described in terms of a conserved quantity, the flux, and its conjugate fractional charge, the triplet.

To clarify this, let us first consider the case ${V_2=V_3=0}$. In this case, Eq.~(\ref{eq:Hamiltonian}) becomes the nearest-neighbor antiferromagnetic Ising model on a triangular lattice. Its ground state is macroscopically degenerate and follows the Coulomb rule~\cite{PhysRev.79.357}. Every triangle must contain either two charges and one vacancy or one charge and two vacancies. Thus, each triangle hosts exactly one frustrated bond, where both sites are occupied by charges or both by vacancies. This manifold includes diagonal and horizontal ordered states, as well as many disordered (isotropic) states [Fig.~\ref{Fig1} (c)].
As a result of this Coulomb rule, we find each row of the triangular lattice supports exactly the same number of frustrated bonds. 
For example, in Fig.~\ref{Fig1}(c) (right), each horizontal row has two frustrated bonds.

The origin of this constraint on frustrated bonds can be seen in an arrow representation~\cite{annurev:/content/journals/10.1146/annurev-conmatphys-070909-104138}, which is a convenient expression to describe the Coulomb phase character of various two-dimensional frustrated magnets~\cite{doi:10.1143/JPSJ.71.2365,PhysRevB.68.064411,PhysRevB.80.140409,PhysRevLett.119.077207}. Placing dual sites on the triangles produces a honeycomb lattice [Fig.~\ref{Fig1} (d)]. On a bond crossing a frustrated bond, we place a vector field (arrow) of magnitude two from a downward triangle to an upward triangle. On all other bonds, we place a vector of magnitude one in the opposite direction. At each dual site, the incoming and outgoing vectors balance each other, thereby satisfying Gauss's law. As a result, for a finite closed region, the incoming and the outgoing arrows must be balanced, which leads to the constant number of frustrated bonds irrespective of the row.
The number of frustrated bonds along each direction, $(L_1, L_2, L_3)$, is conserved under local charge moves preserving the Coulomb rule. We refer to this set of conserved quantities as the flux.

The arrow representation implies the existence of charge as a sink and a source of the vector fields. To see this, starting from a ground state that satisfies the Coulomb rule, a local displacement of a charge can violate this rule,
producing a pair of triangles, one upward, and another downward, on which all the three sites are occupied with charges or vacancies [Fig.~\ref{Fig1} (d)]. We call these triangles as triplets. In the arrow representation, a triplet acts as a source or a sink of arrows: a source on a downward triangle and a sink on an upward triangle.
In the presence of triplets, the flux is no longer conserved. In particular, flux values can be changed by $\pm2$, if the triplets are pair-created, transported around the system, and pair-annihilated [Fig.~\ref{Fig1} (e)].

This conservation law provides the link between flux, triplets, and slow dynamics. The low-temperature ordered phases correspond to anisotropic flux sectors: $(L,0,0)$ for the diagonal order and $(L/2,L/2,0)$ for the horizontal order. In contrast, the high-temperature Coulomb phase belongs mainly to the isotropic flux sector $(L_1,L_2,L_3)\sim(L/3,L/3,L/3)$, which dominates by entropy. When the system is quenched below $T_c$ from the isotropic sector, the change of flux sector is only possible through the motion of triplets. Without this motion, relaxation to the charge-ordered states cannot occur. This requirement imposes a severe kinetic bottleneck and governs the glassy relaxation to the charge-ordered states. 

{\it Method}:
To describe the relaxation dynamics after cooling, we adopt a stochastic dynamics consisting of single charge transfer processes. In each trial, a charge attempts to move to its nearest-neighbor vacant site, and the move is accepted or rejected according to a heat-bath-type transition rate, $W(\Omega_i\to\Omega_f)=\frac{e^{-\beta E_f}}{e^{-\beta E_i} + e^{-\beta E_f}}$, where $E_i (E_f)$ is the system energy before (after) the trial. The Master equation, $\frac{d}{dt}p(\Omega) = \sum_{\Omega'}p(\Omega')W(\Omega'\to\Omega) - p(\Omega)W(\Omega\to\Omega')$, is solved numerically using the kinetic Monte Carlo method, which is applied to the dynamics of frustrated magnetic systems~\cite{PhysRevB.94.104416}.

To characterize the charge order, we define the spatial fraction of the ordered domain, $\Delta_{\rm CO}$, as the number of sites belonging to the largest ordered cluster. When the system is covered by a single domain, $\Delta_{\rm CO}=1$. 

For the cooling protocol, we adopt the zero-triplet quench. The system is first prepared in one of randomly selected ground states of the Coulomb phase at $V_2=V_3=0$, which contains no triplets. At time $t=0$, the temperature is set to $T_{\rm end}$, and
finite interactions $V_2$ and $V_3$ are introduced at the same time. We then study the subsequent dynamics. This protocol enables us to extract the universal features of quench dynamics that originate from the topological-ordered Coulomb phase. In what follows, we focus on quenches into the diagonal phase, with parameters $V_2=0.05$ and $V_3=0$. The system size is $L=144$, unless otherwise noted.

\begin{figure*}[t]
\begin{center}
\includegraphics[width=\textwidth]{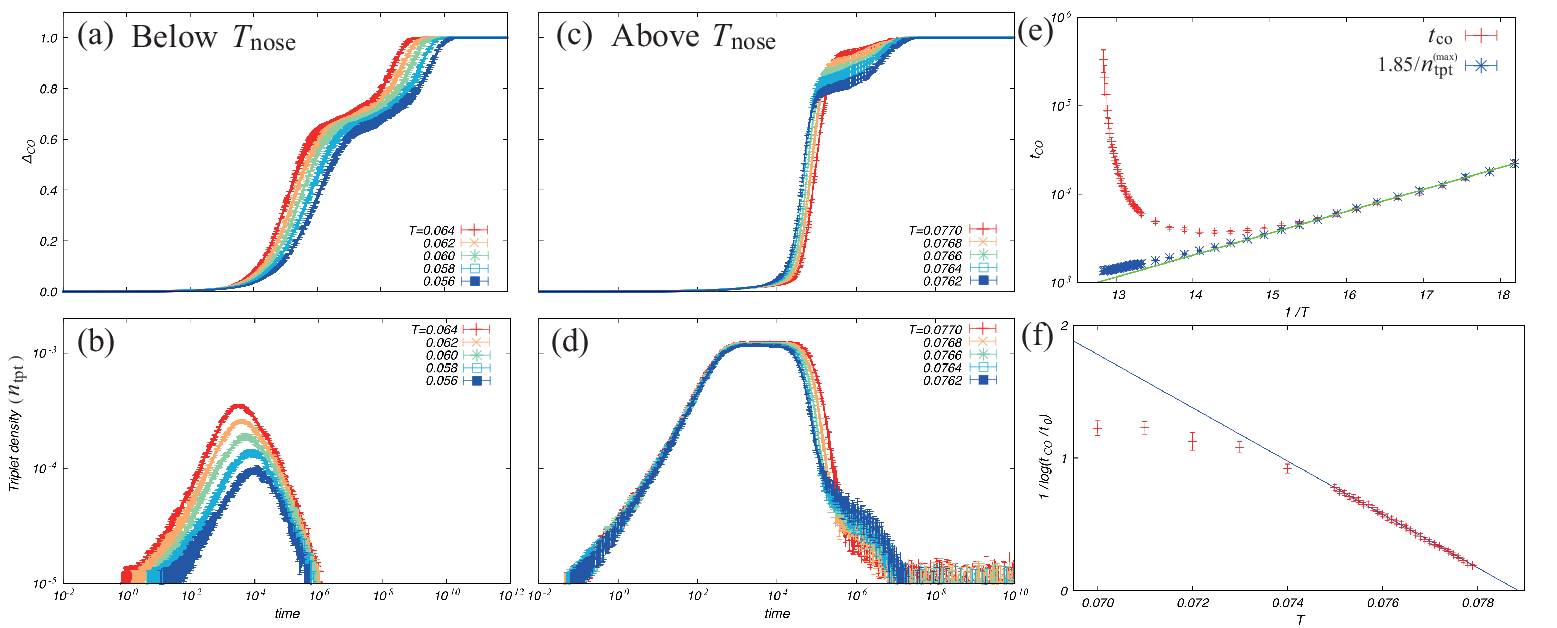}
\end{center}
\caption{\label{Fig2} 
(Color online) Two distinct kinetic regimes of crystallization from a topological ordered liquid. (a,b) Time evolution of the ordered fraction $\Delta_{\rm CO}$ and the triplet density $n_{\rm tpt}$ after quenches to temperatures $T_{\rm end}$ below $T_{\rm nose}$, showing that ordering begins only after $n_{\rm tpt}$ reaches its maximum $n_{\rm tpt}^{\rm (max)}$. (c,d) Corresponding results above $T_{\rm nose}$, where ordering is delayed by a long plateau in $n_{\rm tpt}$, indicating a nucleation bottleneck. (e) Ordering time $t_{\rm CO}$ plotted against inverse temperature $1/T$. At low $T$, $t_{\rm CO}$ is proportional to the inverse of maximal triplet density: $t_{\rm CO}\sim1.85/n_{\rm tpt}^{\rm (max)}$. $n_{\rm tpt}^{\rm (max)}$ follows Arrhenius law, $n_{\rm tpt}^{\rm (max)}\propto\exp(-\frac{0.57V_1}{T})$ as shown with the green solid line. (f) $\log(t_{\rm CO}/t_0)^{-1}$ is plotted against the temperature, $T$. The solid line shows a linear fitting predicted by classical nucleation theory: $\log(t_{\rm CO}/t_0)^{-1}\sim A(T_{\rm c} - T)$ with $t_0=1650$, $A=201.01$, and $T_{\rm c}=0.0789$.
}
\end{figure*}

 {\it TTT diagram}:
 In Fig.~\ref{Fig0} (a), we summarize the time evolution of $\Delta_{\rm CO}$ at each target temperature of quench, $T_{\rm end}$, in TTT diagram. The diagram exhibits a clear nonmonotonic dependence on $T_{\rm end}$. We define the ordering time $t_{\rm CO}$ as the time when $\Delta_{\rm CO}$ reaches 0.05. Then, $t_{\rm CO}$ takes a minimum at $T_{\rm end}=0.071$, which we define as the nose temperature, $T_{\rm nose}$. In the following, we examine separately the two temperature regimes below and above $T_{\rm nose}$.

We first discuss the region below $T_{\rm nose}$. Figure~\ref{Fig2} (a) shows that the onset of charge order is delayed as the temperature decreases. The corresponding triplet density $n_{\rm tpt}$ starts from zero, increases monotonically to a maximal $n_{\rm tpt}^{\rm (max)}$, and then decreases as the charge order develops~[Fig.~\ref{Fig2}~(b)]. Notably, $\Delta_{\rm CO}$ begins to grow only after $n_{\rm tpt}$ reaches its maximum and starts to decrease. As $T_{\rm end}$ increases, the maximum appears earlier, and the onset of ordering shifts to earlier times, accordingly. As shown in Fig.~\ref{Fig2}~(e), the ordering time is inversely proportional to the maximal triplet density, $t_{\rm CO}\sim1.85(n_{\rm tpt}^{\rm (max)})^{-1}$. This scaling provides compelling evidence that the ordering speed is limited by the diluteness of triplets below $T_{\rm nose}$. 

The maximal triplet density follows an Arrhenius law, $n_{\rm tpt}^{\rm (max)} \propto e^{-\Delta/2T}$ [Fig.~\ref{Fig2}(e)], with $\Delta \sim 1.14V_1$, close to the excitation gap of the diagonal charge-ordered state, $V_1+2V_2=1.10V_1$. This behavior differs from the equilibrium distribution, $n_{\rm tpt}^{\rm (eq)} \propto e^{-\Delta/T}$, and instead reflects prethermalization dominated by rapid triplet pair creation and annihilation. A simple rate equation, $\frac{d}{dt}n_{\rm tpt}=\lambda-n_{\rm tpt}^2/\tau$ with $\lambda\propto e^{-\Delta/T}$, yields the steady-state solution $n_{\rm tpt}^{\rm (max)}=\sqrt{\lambda\tau}\propto e^{-\Delta/2T}$, consistent with our results.

Next, we turn to the high temperature region above $T_{\rm nose}$. Here, the ordering time is markedly enhanced compared with what the triplet density alone would suggest~[Fig.~\ref{Fig2}~(e)].
In Fig.~\ref{Fig2}~(c) and (d), we plot evolutions of $\Delta_{\rm CO}$ and $n_{\rm tpt}$ above $T_{\rm nose}$. $\Delta_{\rm CO}$ starts to grow only after a long plateau in the triplet density, which persists up to $t\sim 10000$. This indicates that the presence of sufficient triplets does not immediately trigger ordering, implying a different kinetic bottleneck. 

According to the classical theory of nucleation, the ordered domain continues to grow, only when the domain size exceeds the critical value $R_{\rm c} = \frac{E_{\rm b}}{\Delta E(1-\frac{T}{T_{\rm c}})}$, determined by the balance between the bulk energy gain and the boundary energy cost. Growth of critical nucleus requires overcoming the nucleation barrier of $\Delta_{\rm nuc}=\frac{E_{\rm b}^2}{\Delta E(1-\frac{T}{T_{\rm c}})}$, where $\Delta E$ is the difference of bulk energy density between the ordered and disordered phases, and $E_{\rm b}$ is the domain energy density. The form of $\Delta_{\rm nuc}$ suggests that $\Bigl[\log(t_{\rm CO}/t_0)\Bigr]^{-1}$ behaves as a linear function of the temperature difference from $T_{\rm c}$. Indeed, in the range $0.0750\leq T\leq0.0779$, we find good agreement with a linear fitting: $\Bigl[\log(t_{\rm CO}/t_0)\Bigr]^{-1} = A(T_{\rm c} - T)$, with $t_0=1650$, $A=201.01$ and $T_c=0.0789$ [Fig.~\ref{Fig2} (f)]. This analysis shows that the delay of ordering above $T_{\rm nose}$ is attributed to the time spent on the formation of the critical nucleus.

The ordering kinetics are governed by two competing bottlenecks: rare excitations of triplets at low temperatures and the nucleation barrier at higher temperatures. Their crossover naturally produces the nose temperature observed in the TTT diagram.

\begin{figure}[h]
\begin{center}
\includegraphics[width=0.5\textwidth]{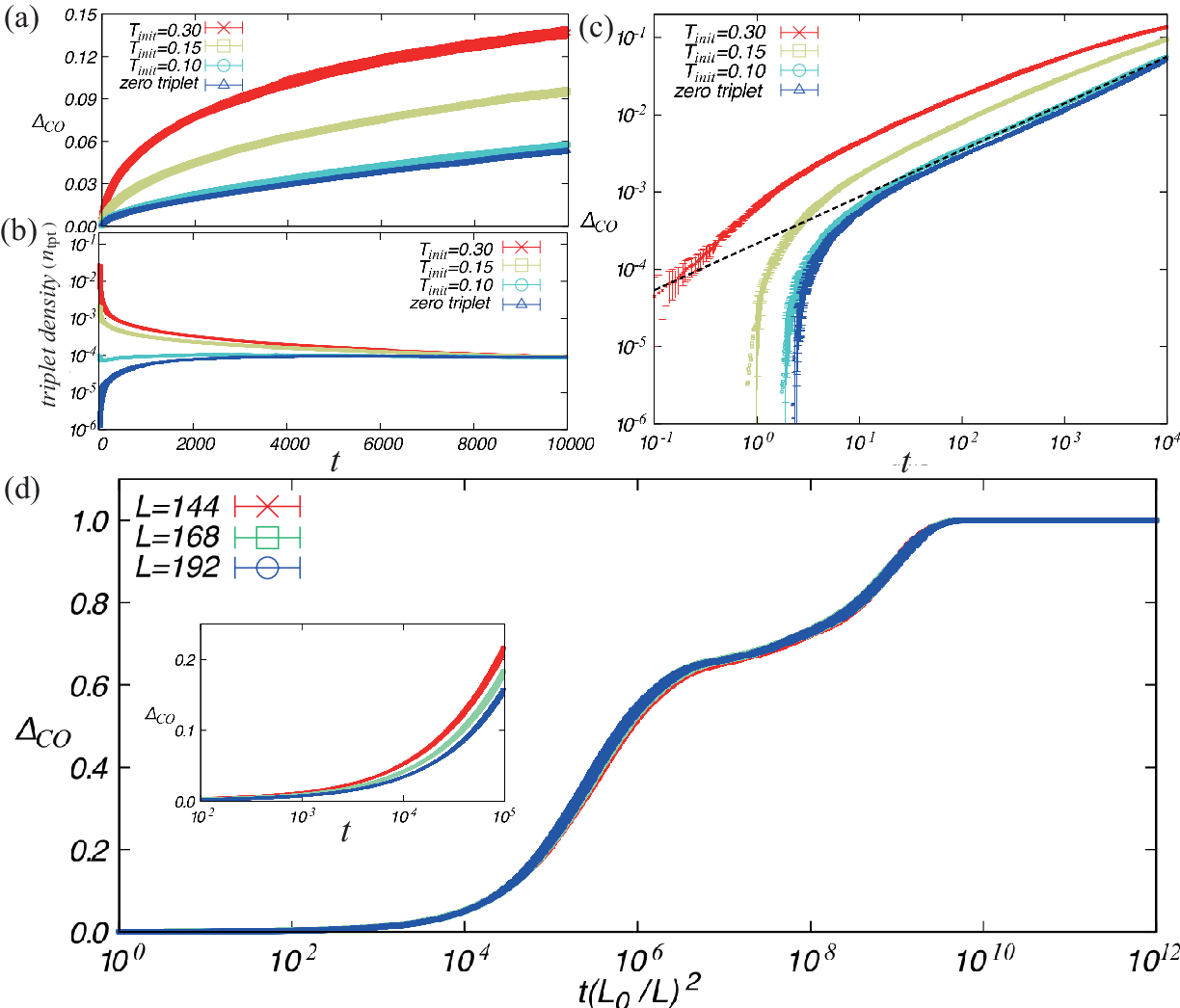}
\end{center}
\caption{\label{Fig3} 
(Color online) Slow ordering process in the initial stage.
(a,b) Time evolution of the ordered fraction $\Delta_{\rm CO}$ and the triplet density $n_{\rm tpt}$ after quenches to $T_{\rm end}=0.06$ from different initial temperatures $T_{\rm init}$ together with the result of the zero-triplet protocol. The growth of $\Delta_{\rm CO}$ accelerates for higher $T_{\rm init}$, reflecting the larger initial triplet population. (c) Log-log plot of (a) with power-law fit $Kt^n$ where $K=2.2\times10^{-4}$ and $n=0.604$ (dashed line), evidencing sublinear growth. (d) Scaling of $\Delta_{\rm CO}$ with the rescaled time $t(L_{0}/L)^{2}$ with $L_0=144$ for the three system sizes, $L=144, 168$, and $192$.The data collapse demonstrates that the ordering time diverges as $L^{2}$. The inset shows $\Delta_{\rm CO}$ in the initial time range, plotted against unscaled time, $t$.
}
\end{figure}

{\it Triplet diffusion}:
Since the small number of triplets causes the dominant dynamical bottleneck below $T_{\rm nose}$, we next examine how their motion affects the early stage of ordering. For this purpose, we consider quenches from finite initial temperatures $T_{\rm init}$, providing a setting closer to experiments, as the early-time dynamics are sensitive to the quench protocol. 

In Fig.~\ref{Fig3}~(a), we plot the evolution of $\Delta_{\rm CO}$ just after a quench to $T_{\rm end}=0.06$, well below $T_{\rm nose}$. The growth of $\Delta_{\rm CO}$ becomes faster for higher $T_{\rm init}$, indicating that the system prepared at a higher temperature orders more rapidly after the quench. In Fig.~\ref{Fig3}~(b), we plot the evolution of the triplet density. In the zero-triplet cooling, the triplet density rises from zero, while in the quench from finite temperature, $T_{\rm init}$, the triplet density takes a finite value already at $t=0$, i.e. the equilibrium value at $T_{\rm init}$, and decreases after the quench [Fig.~\ref{Fig3} (b)]. 

The growth of $\Delta_{\rm CO}(t)$ is initially rapid but soon slows down, exhibiting a sublinear, concave time dependence. The sublinear behavior appears even in the zero-triplet cooling protocol, implying that this slowing down is independent of the initial triplet dynamics. For more quantitative discussions, in Fig.~\ref{Fig3} (c), we show the power-law fits of $\Delta_{\rm CO}(t)=Kt^n$ just after quench. For the quench from $T_{\rm init}=0.1$, in which the triplet density is almost constant, $\Delta_{\rm CO}(t)$ is well fitted with the parameter $n\sim0.604$, close to the Avrami exponent $n\sim0.56$ observed for TlZn system.
This is in sharp contrast to a typical crystallization process, where growth occurs on the surfaces of expanding nuclei, leading to an exponent of $n=3$ in a two-dimensional system, implying that growth accelerates as ordering proceeds. 

The origin of this initial slowing down can be attributed to the dynamics of triplets. As illustrated in Fig.~\ref{Fig1}~(e), the flux changes by $\pm2$ when a triplet circulates around the system and pair-annihilates. Starting from the isotropic flux sector, triplets need to make $\mathcal{O}(L)$ circulations in total to achieve full ordering. Suppose that a triplet makes a normal diffusive motion with diffusion constant $D$. The time required to circulate around the system is then $t^*=\frac{L^2}{D}$. The total number of circulations up to time $t$ is $N_{\rm tpt}\frac{t}{t^*}$, where $N_{\rm tpt}$ is the number of triplets. Because the direction of triplet motion is random, its fluctuation $\sqrt{N_{\rm tpt}\frac{t}{t^*}}$ effectively contributes to the ordering [Fig.~\ref{Fig1} (e)]. The number of triplets is proportional to the system size $L^2$ and suppressed by the Arrhenius factor, $e^{-\frac{\Delta}{2T}}$. Consequently, the ordering domain evolves as
\begin{eqnarray}
\Delta_{\rm CO}\propto\frac{1}{L}\sqrt{N_{\rm tpt}\frac{t}{t^*}}\propto e^{-\frac{\Delta}{4T}}\sqrt{\frac{Dt}{L^2}}.
\label{eq:tripletdiffusion}
\end{eqnarray}
This relation gives the exponent $n=0.5$, which is close to the numerical value $n\sim0.604$.

Equation~(\ref{eq:tripletdiffusion}) further implies a distinct system-size dependence of the dynamics. Because $\Delta_{\rm CO}$ scales with the single parameter $\frac{Dt}{L^2}$, the ordering time diverges in the thermodynamic limit $L\to\infty$. Figure~\ref{Fig3}~(d) shows the evolution of $\Delta_{\rm CO}$ at $T_{\rm end}=0.06$ plotted against the rescaled time $t(\frac{L_0}{L})^2$ for several system sizes. The data collapse onto a single curve confirms the scaling relation Eq.~(\ref{eq:tripletdiffusion}). Intuitively, this scaling means that triplets must diffuse over the entire system to change the value of fluxes and establish charge ordering. This ``crystallization by diffusion" mechanism results in macroscopic relaxation time and glassy behavior.

 {\it Discussions}: We have studied the charge-ordering dynamics of the supercooled $V_1$-$V_2$-$V_3$ charge Ising model. The model possesses the topological-ordered liquid phase characterized by a global conserved quantity, called flux, and elementary excitations called triplets. We found that triplets play a crucial role in the ordering dynamics. At low target temperatures $T_{\rm end}$, the ordering rate is proportional to the triplet density, leading to an Arrhenius law-like increase of the ordering time. 
At higher temperatures, in contrast, the nucleation barrier dominates.
The competition between those two mechanisms naturally leads to the nose temperature in the TTT diagram. Moreover, the ordering requires a change in flux sectors, which can occur only through triplet diffusion across the entire system, leading to a distinct system size dependence of the ordering timescale and anomalously small Avrami exponent. 
 
These characteristics of the supercooled topological-ordered liquid well explain the main features of the charge glass phenomenon observed in $\theta$-(BEDT-TTF)$_2$X(SCN)$_4$; the origin of the glassy behavior, the initial slowing down of the ordering dynamics, and the structure of TTT diagram with the nose temperature. We expect triplet excitations to play a key role in charge glass behavior, and its direct experimental observation is strongly desired. Detection of triplets is not straightforward, since the triplets are charge-neutral on average: their electric charges rapidly fluctuate. Nevertheless, presumably, the optical conductivity measurements already caught a glimpse through the peaks in the spectra~\cite{PhysRevB.89.085107,cryst12060831}. It is interesting to clarify the relation between the optical peaks with the energy scales of triplets, and obtain further insights into their dynamical characters.

It is also interesting to point out that the coexistence of long-period charge fluctuations is reported for X$=$TlCo, RbZn and CsZn materials~\cite{doi:10.1143/JPSJ.79.044606,PhysRevLett.93.216405,doi:10.1143/JPSJ.68.2654, Sawano:2005aa,doi:10.7566/JPSJ.83.083602,doi:10.7566/JPSJ.83.083602,PhysRevB.89.121102}, which may be associated with nonlinear I-V characteristics~\cite{doi:10.1143/JPSJ.78.024714,Sawano:2005aa,PhysRevLett.96.136602}. The spatial periodicity of these fluctuations is different from the charge ordering patterns in the ordered phase. Even though the charge order is suppressed by rapid cooling, these long-period fluctuations persist. From the perspective of global fluxes, it is tempting to raise the possibility that the long-period fluctuation may be the charge instability specific to the isotropic flux sector, possibly connected with the lattice instability~\cite{PhysRevB.105.L041114,PhysRevResearch.6.023003}. If the system fails to relax from the high-temperature isotropic flux sector, the system may try to reorganize itself into an optimal charge configuration within the flux sector.

Our findings represent universal features of supercooling from topological-ordered liquid. Glassy dynamics has been observed for a broad range of frustrated systems, and in most cases, they are vaguely attributed to the complexity of the system due to frustration. Our scenario may give an illuminating viewpoint on some of the unresolved issues. In this light, it is interesting to point out the possible relevance of our conclusion to water ice, a typical system with Coulomb phase character~\cite{Li01061994,PhysRevB.91.245152}. Needless to say, water is the most important substance for us. We wish the notion of supercooling topological-ordered liquid will be useful in clarifying still abundant mysteries of water ice~\cite{Bartels-Rausch:2013aa}.

 {\it Acknowledgements}: 
This work was supported by the JSPS KAKENHI (Nos. JP20H05655, JP22H01147, JP23K22418, 	JP23K13031, and JP25H01401), MEXT, Japan.
The authors are grateful to K. Hashimoto for insightful discussions and for sharing essential information on Avrami exponent observed for X$=$TlZn compounds.
We also thank K. Kanoda, H. Oike, and M. Kato for useful discussions.

\bibliographystyle{apsrev4-1}
%


\widetext
\pagebreak

\renewcommand{\theequation}{S\arabic{equation}}
\renewcommand{\thefigure}{S\arabic{figure}}
\renewcommand{\thetable}{S\arabic{table}}
\setcounter{equation}{0}
\setcounter{figure}{0}
\setcounter{table}{0}

\begin{center}
\Large 
{Supplemental Material}
\end{center}

%
\subsection{Kinetic Monte Carlo simulation}
Kinetic Monte Carlo (KMC) simulation is a numerical method to solve stochastic equations, which normally takes the form of
\begin{eqnarray}
\frac{d}{dt}P(\Omega, t) = \sum_{\Omega'}P(\Omega', t)W(\Omega'\to\Omega) - P(\Omega, t)W(\Omega\to\Omega').
\end{eqnarray}
In our context, $\Omega$ stands for a charge configuration, and $P(\Omega, t)$ is the probability that the configuration $\Omega$ is realized at time $t$.
$W(\Omega\to\Omega')$ is the transition rate from $\Omega$ and $\Omega'$.
We assume the transition is possible, if $\Omega'$ is connected with $\Omega$ by a single charge transfer, and adopt the thermal-bath-type form, 
\begin{eqnarray}
W(\Omega\to\Omega') = \frac{e^{-\beta E_{\Omega'}}}{e^{-\beta E_{\Omega}} + e^{-\beta E_{\Omega'}}},
\end{eqnarray}
with inverse temperature, $\beta=\frac{1}{k_{\rm B}T}$, and the system energy $E_{\Omega}$.
This choice of transition rate ensures that the system relaxes to the equilibrium state described by the Boltzmann distribution in the long time limit.

\begin{figure}[h]
\begin{center}
\includegraphics[width=0.7\textwidth]{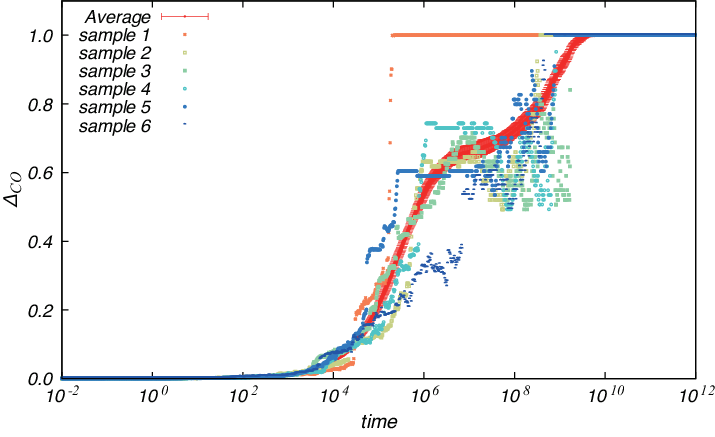}
\end{center}
\caption{\label{Fig_S1} 
(Color online) Time evolution of the order fraction $\Delta_{\rm CO}$ obtained by KMC method for the zero-triplet quench to $T_{\rm end}=0.06$. Time series for 6 different samples, and the average over 1000 samples are shown.}
\end{figure}

The procedure of KMC method is divided into several parts. Suppose that charge configuration $\Omega^{(n)}$ is realized at the $n$-th step of the simulation and the time is set to $t_n$. Then, one needs to do the following:
\begin{itemize}
\item[1.] list up all the charge configurations $\{\Omega_m|m=1, \cdots N_{\Omega^{(n)}}\}$ accessible from $\Omega^{(n)}$
\item[2.] Calculate all the transition rate $W(\Omega^{(n)}\to\Omega_m)$ and their total sum, $W_{\rm tot}\equiv\sum_{m=1}^{N_{\Omega^{(n)}}}W(\Omega^{(n)}\to\Omega_m)$
\item[3.] use the random number $r_1\in[0,1]$ to choose the process $m_0$, which actually occurs, according to the probability, $\frac{W(\Omega^{(n)}\to\Omega_{m_0})}{W_{\rm tot}}$ 
\item[4.] use the second random number $r_2\in[0,1]$ to determine the time $\Delta t$ spent by the transition: $\Omega^{(n)}\to\Omega_{m_0}$, according to $\Delta t=-\frac{1}{W_{\rm tot}}\log r_2$
\item[5.] update the time: $t_n\to t_{n+1}=t_n+\Delta t$, and the configuration, $\Omega^{(n)}\to\Omega^{(n+1)}=\Omega_{m_0}$, and return to 1.
\end{itemize}
For the evaluation of physical quantities, such as the ordered fraction $\Delta_{\rm CO}$ and the triplet density $n_{\rm tpt}$, we set the time series for observation in advance, e.g. we set
$\{t_{\rm obs}^{i} = \frac{i}{N_{\rm obs}}T|i=0, \cdots N_{\rm obs}\}$, or for a long time simulation, we use logarithmic discretization, $\{t_{\rm obs}^{i} = 10^{n_{\rm init} + i\Delta n}|i=0, \cdots N_{\rm obs}\}$.
For all the observation times $t_{\rm obs}^{i}$ included in the $n$-th time range, $t_n < t^i_{\rm obs} < t_{n+1}$, we assign the same physical quantities evaluated at the time $t_n$.
Accordingly, from one simulation, we obtain a time series of physical quantities in a form of step function.
We make sample average and obtain a smooth curve [Fig.~\ref{Fig_S1}].
We typically perform the calculations for the system of $L=144$ (20736 sites), unless specially noted, and make average over 1000 samples for each target temperature, $T_{\rm end}$. 

\subsection{Derivation of critical radius}
The transition to the diagonal charge ordered phase is a strong first-order phase transition.
Accordingly, if the system is quenched to just below the transition temperature, $T_{\rm c}$, an ordered domain grows after a sufficiently large nucleus is generated, according to the standard theory of phase transition dynamics.
\begin{figure}[h]
\begin{center}
\includegraphics[width=0.6\textwidth]{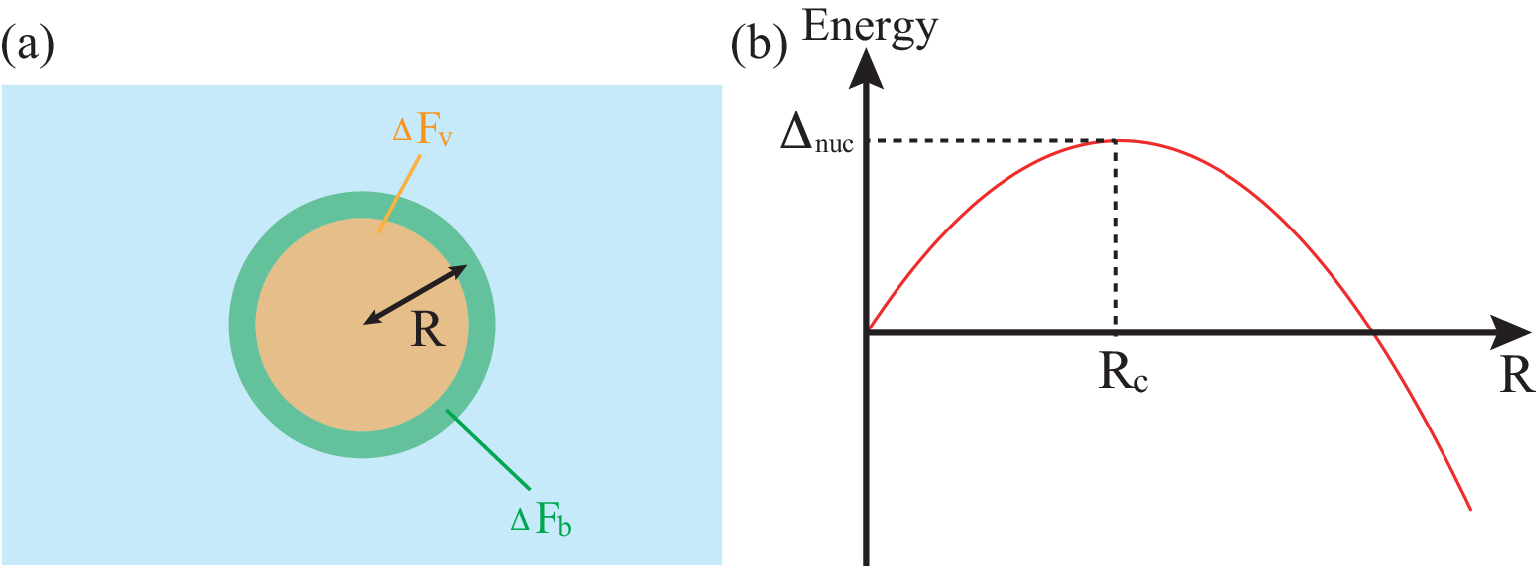}
\end{center}
\caption{\label{Fig_S2} 
(Color online) (a) Schematic figure of a nucleus of the ordered domain of radius $R$. (b) Schematic figure of the energy cost, $\Delta_{\rm nuc}$, by creating a nucleus of radius, $R$.}
\end{figure}
The critical nucleation radius $R_{\rm c}$ can be estimated as follows.
Suppose the system is set at the temperature $T$, which is slightly lower than the transition temperature, $T_{\rm CO}$.
Then, the bulk free energy gain by nucleating a circular charge ordered domain of radius $R$ can be estimated as
\begin{align*}
\Delta F_{\rm v} = \pi R^2(F_{\rm CO}-F_{\rm normal}) = -\pi R^2\Delta S(T_{\rm c} - T) < 0
\end{align*}
Meanwhile, the boundary costs the free energy,
\vspace{-0.5cm}
\begin{align*}
\Delta F_{\rm b} = 2\pi RE_{\rm b} > 0
\end{align*}
Critical nucleation radius ($R_{\rm c}$) and the associated energy barrier ($\Delta_{\rm nuc}$) is determined by minimizing $\Delta F_{\rm v} + \Delta F_{\rm b}$, and is obtained as
\begin{eqnarray*}
\begin{cases}
R_{\rm c} = \frac{E_{\rm b}}{\Delta S(T_{\rm c} - T)},\\
\Delta_{\rm nuc} = \pi\frac{E_{\rm b}^2}{\Delta S(T_{\rm c} - T)},
\end{cases}
\end{eqnarray*}
which leads to the time necessary to make the ordered domain larger than critical radius, as $\displaystyle t_{\rm nuc}\propto e^{\frac{\Delta_{\rm nuc}}{k_{\rm B}T_{\rm c}}}\sim e^{\frac{E_{\rm b}^2}{\Delta E(1 - \frac{T}{T_{\rm c}})}}$.
This relation can be transformed into
\begin{eqnarray}
\frac{1}{\log(t_{\rm CO}/t_0)} = A(T_{\rm c} - T),
\label{eq:CriticalNucleationTheory}
\end{eqnarray}
with three fitting parameters, $t_0, A$, and $T_{\rm c}$.

Eq. (\ref{eq:CriticalNucleationTheory}) is used to fit $t_{\rm CO}$ above $T_{\rm nose}$, as shown in Fig.~3 (f) in the main text.
To fit the data, we sweep $t_0$, and at each fixed value of $t_0$, we make the least square fitting to $\frac{1}{\log(t_{\rm CO}/t_0)}$ with the two parameters, $A$ and $T_{\rm c}$ in the temperature range, $0.0750\leq T\leq0.0779$, and evaluate the mean squared error, $s^2$. Then, from the minimum of $s^2$, we determine $t_0=1650$, and find $A=201.01$ and $T_{\rm c}=0.0789$ as corresponding values.

\subsection{Identification of ordered regions}
\begin{figure}[h]
\begin{center}
\includegraphics[width=0.8\textwidth]{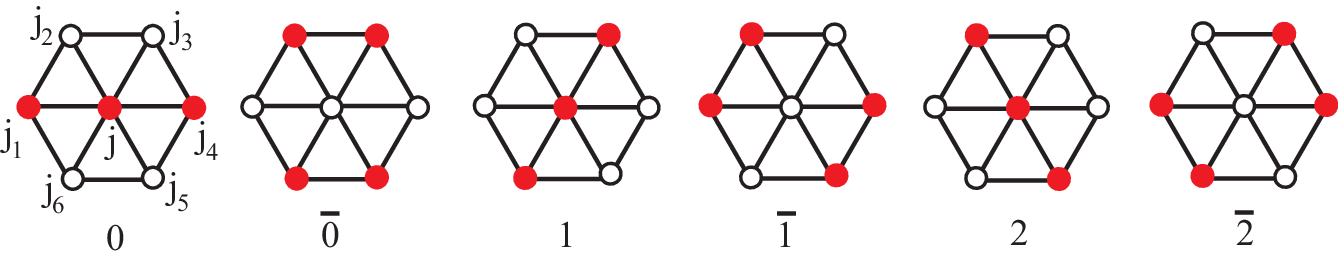}
\end{center}
\caption{\label{Fig_S3} 
(Color online) 6 different types of ordering units. In the left figure, the site index $j$, and the indices of surrounding sites, $j_1\cdots j_6$, are shown.}
\end{figure}

To evaluate the ordered fraction, $\Delta_{\rm CO}$, we first identify ordered domains in the real space. To distinguish three different orientational patterns of diagonal charge order, we first introduce local ordering units as shown in Fig.~\ref{Fig_S3}.
Each ordering unit is defined for a hexagon centered around a site. 
Suppose a local charge configuration around a site $j$ matches e.g. the ordering unit $\bar{1}$, we assign the local order parameter $\bar{1}$ to the site $j$.
If none of the ordering units matches the local charge configuration, we assign ``none".

After assigning an ordering unit to each site, we identify ordered domains one by one in the depth-first search protocol.
Suppose we start with the site $0$, and increase the site index, and the assigned ordering unit is not ``none" for the first time at site $j$.
Then, we register the site $j$ as the first member of the first domain.
Suppose that the ordering unit $0$ is assigned to the site $j$, as shown in the left panel of Fig.~\ref{Fig_S3}.
After identifying the first member of the domain, $j$, we move to its neighboring sites; $j_1,\cdots, j_6$ [Fig.~\ref{Fig_S3}, left].
If the ordering unit 0 covers the system, the site $j_1$ should have the ordering unit $0$.
So, if the site $j_1$ is actually assigned the ordering unit $0$, we add the site $j_1$ as the next member of the first domain.
Similarly, if the site $j_2$ has the ordering unit $\bar{0}$ (not $0$), then the site $j_2$ belongs to the same domain, so we add the site $j_2$ as a new member.
From the added members among $j_1,\cdots, j_6$, we extend search to their neighboring sites and find all the members of the first domain.
Then, we move to the search for a next domain, and classify all the sites in the system into the domains, and the sites with ``none".

\end{document}